\documentclass[twocolumn,nofootinbib,prd,preprintnumbers,amsmath,amssymb]{revtex4}
\usepackage[hypertex]{hyperref}
\usepackage{graphicx}
\usepackage{epsfig}
\newcommand{\be}{\begin{equation}}
\newcommand{\ee}{\end{equation}}
\newcommand{\ba}{\begin{eqnarray}}
\newcommand{\ea}{\end{eqnarray}}

\begin{document}
\preprint{} 
\title{SUSY Digs up a Buried Higgs}
\author{Brando Bellazzini$^{a}$}
\author{Csaba Cs\'aki$^{a}$}
\author{Jay Hubisz$^{b}$}
\author{Jing Shao$^{b}$}
\affiliation{
\vspace{0.3cm} $^a$	Institute for High Energy Phenomenology,
Newman Laboratory of Elementary Particle Physics,Cornell University, Ithaca, NY 14853, USA \\ \\
$^b$201 Physics Building, Syracuse University, Syracuse, NY 13244, USA}
\date{\today}

\begin{abstract}

The Higgs boson may dominantly decay to $4$ light jets through a light pseudo-scalar intermediary:  $h \rightarrow 2 \eta \rightarrow 4j$, making reconstruction at the LHC particularly challenging.  We explore the phenomenology of  such ``Buried Higgs" scenarios in which the primary discovery channel of the Higgs is in cascade decays of superpartners.  QCD backgrounds that would otherwise overwhelm the Higgs decay are suppressed by the requirement of high $p_T$ jets and large missing transverse momentum that are the typical signatures of TeV scale supersymmetry.  Utilizing jet substructure techniques, we find that for buried Higgses in the $100-120$ GeV range, a 5$\sigma$ discovery can be expected with roughly $10-25$~fb$^{-1}$ of data at $E_\text{CM} = 14$~TeV.  For lighter Higgs bosons, the signal is contaminated by hadronically decaying $W$ bosons, and discovery remains an unsolved challenge.

\end{abstract}

\maketitle

\section{Introduction}

Understanding the mechanism of electroweak symmetry breaking (EWSB) is the primary  
 goal of the  Large Hadron Collider (LHC) program. 
 Within the Standard Model (SM) of particle physics, EWSB occurs via the Higgs mechanism which predicts a weak scale Higgs boson $h$ which unitarizes longitudinal $W$ boson scattering 
  and accommodates electroweak (EW) precision data.
The current limits on the SM Higgs boson mass $m_{h}$ consist of the lower bound $m_{h}\gtrsim 114$ GeV  from LEPII \cite{Barate:2003sz}, and the exclusion window $158<m_{h}<175$ GeV from Tevatron \cite{:2010ar}. 
However, these limits are model dependent. 
For instance, in several extensions of the SM with an extended Higgs sector, new decay channels for the Higgs boson are available and the exclusion bounds are weakened \cite{Chang:2008cw,Dermisek:2005ar,Dobrescu:2000yn,Carena:2002bb}.  Therefore, new strategies to discover non-standard Higgs bosons at the LHC and the Tevatron need to be developed to be sure all regions of EWSB parameter space are covered.

In this work we consider the possibility that a light Higgs boson is produced in supersymmetric cascades and subsequently decays to four light jets (i.e without $b$'s), $h\rightarrow2\eta\rightarrow 4j$, where $\eta$ is a new light pseudoscalar 
that is neutral under the SM gauge group.
This non-standard Higgs boson decay is present in many extensions of the SM  that stabilize the EW scale
\cite{Chang:2005ht,Chang:2008cw,Gripaios:2009pe, Bellazzini:2009xt,Bellazzini:2009kw,Carpenter:2008sy,Carpenter:2007zz}.
The typical fine-tuning in these models  is reduced because the experimental constraints  on the Higgs sector are much weaker. 
Indeed, for $\eta$ heavier than $12$ GeV only the OPAL model independent bound \cite{Abbiendi:2002qp}, $m_{h}>82$ GeV, holds. 
If $m_\eta<12$ GeV,  the lower bound is slightly stronger \cite{Abbiendi:2002in}, $m_{h}>86$ GeV, 
while recently recasted \cite{Cranmer:2010hk} ALEPH data \cite{Schael:2010aw} set $m_\eta\gtrsim 4$ GeV for $m_h=100$ GeV.
Present hadron collider data do not constrain such a Higgs boson because of the huge QCD background.

Recent studies \cite{Falkowski:2010hi,Chen:2010wk} have discussed search strategies 
of a Higgs boson decaying to four light jets at the LHC by means of jets substructure techniques \cite{Butterworth:2008iy}. \footnote{For an alternative search based on tagged outgoing forward protons see \cite{Khoze:2010ba}.}
If the Higgs boson has a large boost factor, its decay products will be merged into a single fat jet with characteristic substructure.  Since QCD jets from background processes are relatively uncorrelated, substructure requirements can significantly enhance the signal to background.
Our analysis extends previous works in three main directions:  
\begin{itemize}
\item
We consider Higgs boson production in association  
with supersymmetric particles which can lead to a much earlier discovery (similar to the $h\rightarrow b\bar{b}$ analysis in SUSY \cite{Kribs:2009yh,Kribs:2010hp}) compared to the search using SM-like production.  In particular, we consider neutralino and chargino two body decays, 
$\chi^0_{i}\rightarrow \chi^0_{j} h$ and $\chi^\pm_{i}\rightarrow \chi^{\pm}_{j} h$.
\item
We do not restrict our analysis to very light $\eta$ below the $b\bar{b}$ threshold.  
We instead extend the $\eta$ mass range,  $m_{\eta}\lesssim 30$~GeV. 
\item
We discuss how to reduce the contamination from hadronic $W$ and $Z$ for a low-mass Higgs boson. 
These are the irreducible backgrounds for supersymmetric Higgs productions and
can be larger than the signal in the resonance regions. This requires special treatment 
because they can easily produce a similar substructure pattern as the Higgs boson.
\end{itemize}

We find that for $m_{h}\lesssim 95$ GeV 
and for a light pseudo-scalar $m_\eta =10$ GeV, the Higgs boson remains hidden in the W background after all cuts are applied.  For heavier Higgs bosons, where the $W$'s are well separated from the Higgs boson peak, we obtain $5\sigma$ signal significance at the LHC with $\sqrt{s}=14$ TeV and $10-25$~$\mbox{fb}^{-1}$ which is smaller by a factor $5-10$ than the luminosity needed for discovery in the SM production channel~\cite{Falkowski:2010hi,Chen:2010wk}.

This paper is organized as follows.
In Sec.~\ref{model}, we discuss some illustrative models in which the Higgs originates in a supersymmetry cascade and subsequently decays to four light jets.
In Sec.~\ref{production}, we discuss rates of Higgs boson production associated with supersymmetric particles. These 
supersymmetric events are isolated by means of cuts discussed in Sec.~\ref{SUSYcuts}.
In Sec.~\ref{substructure} we discuss our analysis, including a refinement of jet substructure algorithms used in our search, 
and we present the results in Section~\ref{results}. 
Finally, we conclude in Sec.~\ref{conclusions}.

\section{Higgs decay to four jets}
\label{model}

Cascade decays of the Higgs boson to SM states through a pair of light pseudoscalar $\eta$'s 
are well motivated in several extensions of the SM.
They include supersymmetric \cite{Chang:2005ht,Chang:2008cw,Bellazzini:2009xt,Bellazzini:2009kw,Carpenter:2008sy,Carpenter:2007zz}
as well as non-supersymmetic \cite{Gripaios:2009pe} realizations. 
After spontaneous EWSB, a trilinear coupling between light SM singlets $\eta$ and the Higgs boson $h$ is a generic feature:
\begin{equation}
\label{h2eta}
\mathcal{L}_{h\eta\eta}\simeq a_{h\eta\eta}\frac{m_h^2}{2f} h\eta^2
\end{equation}
where $f$ is the typical scale that controls the interaction strength.
For instance, in models where $h$ and $\eta$ emerge as pseudo-Goldstone bosons
and $m_{\eta}\ll m_{h}$, the derivative interaction $h(\partial_{\mu}\eta)^2 a_{h\eta\eta}/f $ produces the Lagrangian (\ref{h2eta}) 
after integration by parts. In these scenarios $f$ is analogous to the pion decay constant 
that sets the couplings in the chiral Lagrangian.

If $f$ is not too large compared to the EW scale and $a_{h\eta\eta}$ is not too small, 
the decay width of the Higgs boson into two $\eta$'s 
\begin{equation}
\Gamma_{h\rightarrow\eta\eta}\simeq \frac{a_{h\eta\eta}^2}{32\pi} \frac{m_h^3}{f^2} \sqrt{1-\frac{4m_{\eta}^2}{m_{h}^2}}
\end{equation}
can easily dominate over the SM channel $h\rightarrow b\bar{b}$.
The pseudoscalar $\eta$ is generically unstable because it couples to SM fermions  with
effective Yukawa interactions $i\tilde{y}_{\psi} \eta\bar{\psi}\gamma_{5} \psi$
generated after EWSB, $\tilde{y}_{\psi}\approx m_{\psi}/\sqrt{2}f$. 
Then, barring accidental cancellations, the largest Yukawa coupling is to the third generation quarks, 
while all other SM fermion couplings are highly suppressed. For recent studies where $\eta\rightarrow b\bar{b}$ see Ref.~\cite{Cheung:2007sva,Carena:2007jk,Kaplan:2009qt}.
However, when $m_{\eta}\lesssim 2m_{b}\sim 10$ GeV the decay of $\eta$ to two gluons via loops of third generation quarks will be the dominant decay mode, corresponding to a four unflavored jet final state, $h\rightarrow 2\eta\rightarrow 4g$. 
For instance, this scenario is naturally realized in the supersymmetric ``Buried Higgs'' model \cite{Bellazzini:2009xt}
where both $h$ and $\eta$ are pseudo-Goldstone bosons arising from a global $SU(3)_{H}$ 
symmetry broken down to $SU(2)_{H}$ at the scale $f\approx 500$ GeV. 
In this model, the coupling between $\eta$ and $h$ 
depends only on a mixing angle $v/f$ which measures the alignment between the gauged $SU(2)_{W}$ and the residual global $SU(2)_{H}$,  $a_{h\eta\eta}\approx v/(\sqrt{2}f)(1-v^2/f^2)^{-1/2}$.  The branching ratios for $h\rightarrow 2\eta$ and $\eta\rightarrow 2g$ are $80-90\%$ and $100\%$  respectively.

An appealing modification of the Buried Higgs scenario is the ``Charming Higgs'' model \cite{Bellazzini:2009kw}.
The Higgs sector is the same as in the original Buried Higgs model but the embedding 
of the matter content into the $SU(3)_{H}$ global symmetry
 multiplets is different.  In particular, the bottom Yukawa arises only from non-renormalizable operators  suppressed by the physical scale $\Lambda\approx 10$ TeV where new heavy states are integrated out. 
 Then, the resulting bottom Yukawa coupling to $\eta$ is greatly suppressed because 
 it has to vanish both for large $f$ and for large $\Lambda$, therefore $\tilde{y}_{b}\simeq m_{b}/\sqrt{2}f \times m_{b}^2/\Lambda^2\ll1$ .   Thus, it turns out that the dominant decay channel is the tree-level 
 $\eta\rightarrow c\bar{c}$ even when $\eta$ is above $b\bar{b}$ threshold production, $m_\eta>2m_{b}$.
The next relevant decay mode, $\eta\rightarrow 2g$, is generated  at 1-loop. Very much like the original Buried Higgs model, the charming version buries the Higgs boson beneath the QCD background at the LHC.

Another class of models where the Higgs may naturally cascade decay to four jets is provided by
the non-minimal composite Higgs models \cite{Gripaios:2009pe}.

In the following, we will use the Buried/Charming Higgs models as illustrative examples of supersymmetric extensions of the SM where the Higgs boson cascade decays to four light jets where $m_{\eta}\in[5-30]$ GeV and $m_{h}\in[90-120]$ GeV. 
For simplicity and clarity of presentation, in our simulations we consider Higgs boson production rates as they appear in the Minimal Supersymmetric Standard Model (MSSM). \footnote{Up to $v^2/f^2$ corrections, these rates are the same as those in the buried and charming Higgs models.}
In particular the Higgs production from SUSY events is matched to the production in the MSSM with the same input parameters and Higgs mass. In our numerical study we take the branching ratios for $h\rightarrow 2\eta$ and $\eta\rightarrow 2j$ to both be $100 \%$.

\section{Higgs Production}
\label{production}

In general, there are several possible production channels for the Higgs boson in SUSY in addition to the SM channels. 
The important ones are those with large cross sections such as the pair production of gluinos and squarks. 
In the subsequent cascade decay of these particles, a Higgs boson can be produced in many different ways, 
the most important being through supersymmetric gauge interactions from the gaugino-Higgsino-Higgs coupling. 
Depending on the mass spectra of charginos and neutralinos, there can be two scenarios: the ``little cascade"  and the ``big cascade".

In the first case, a Higgs is produced in the decay $\chi_2\rightarrow \chi_1 h$.  For this channel to dominate, the $\mu$-term needs to be larger than the gaugino masses to ensure that $\chi_{1}^{\pm}$ and $\chi_{1,2}^{0}$ are gaugino-like.  If, in addition, the sleptons are heavier than $\chi_2^{0}$, the decay to Higgs is always dominant since the other available mode,  $\chi_2^0\rightarrow\chi_1^0 Z$ depends on the Higgsino-Higgsino-Z coupling which is doubly suppressed by small Higgsino mixing.  The $\chi_2$ can be produced from the decays of $\tilde q$ or $\tilde g$ with sizable branching fractions, producing a sizable fraction of SUSY events which contain Higgs bosons. 
 
In the second case, the Higgs boson is produced in the decay of heavier neutralinos and charginos ($\chi_{2}^{\pm}$/$\chi_{3,4}^{0}$) into lighter ones ($\chi_{1}^{\pm}$/$\chi_{1,2}^{0}$). To have a large rate from gluino and squark decays, the heavy neutralinos and charginos need to be gaugino-like whereas the light ones Higgsino-like. This corresponds to a $\mu$-term which is small in comparison with the gaugino masses, and to weak gauginos which are lighter than squarks and gluinos. 
 
\begin{figure}[htbp]
\begin{center}   
   \includegraphics[scale=0.7]{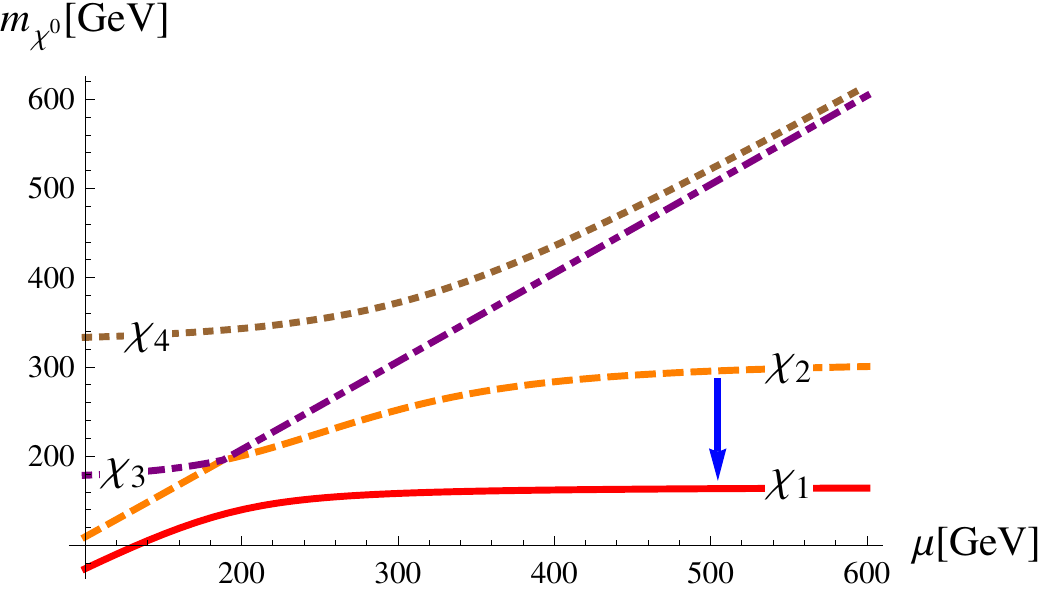}
   	\includegraphics[scale=0.7]{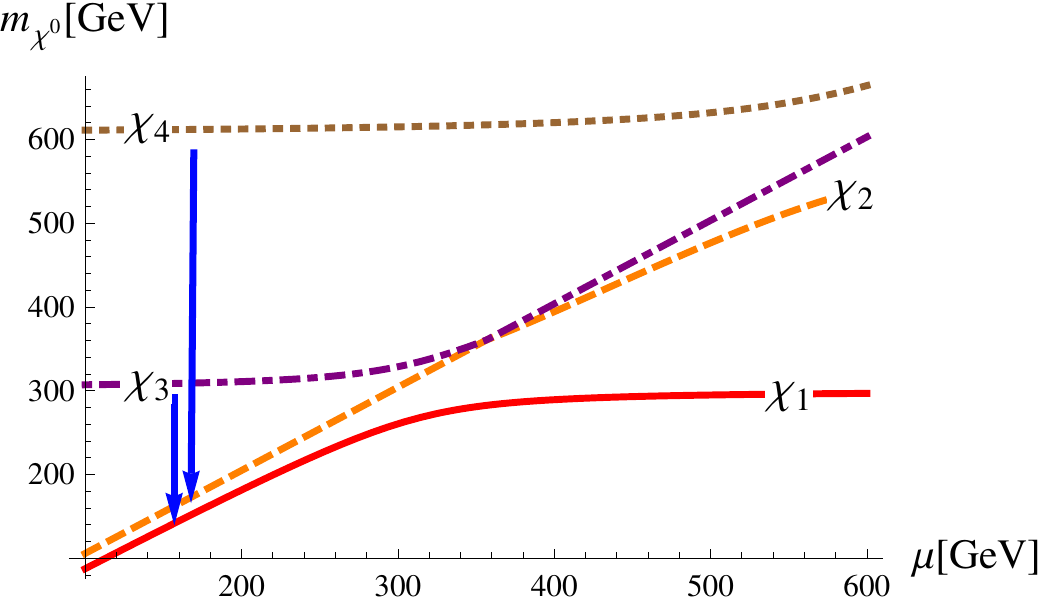}
\caption{
 Top: Neutralino mass spectrum in the little cascade scenario where $\chi^0_{1,2}$ are non-degenerate and gaugino like for large $\mu$.
 Bottom: Neutralino mass spectrum in the big cascade scenario where $\chi^0_{1,2}$ are almost degenerate and Higgsino like for small $\mu$. 
 Blue arrows show the decay mode relevant for the present search.}
\label{neutralino-masses}
\end{center}
\end{figure}

Fig.~\ref{neutralino-masses} shows the typical netralino mass spectrum for the ``big" and ``little" cascade scenarios.

In our analyses, we consider two benchmark SUSY models with parameters given in Tab.~\ref{table:mass}. 
Benchmark 1 is an mSUGRA model with $m_{1/2}=m_0=400$~GeV, $A_0=0$, $\tan\beta=10$ and $\text{sgn}(\mu)=1$. 
It corresponds to a model where the Higgs is produced in the little cascade  $\chi_{2}\rightarrow \chi_{1}h$. The dominant production rate is from the pair production of gluinos/squarks, which is given in Tab.~\ref{table:mass}. The generic final states are $ (\ge 2) q + 2\left[W/Z/h\right]+ E_T^{\text{miss}}$. As can be seen from Tab.~\ref{table:mass}, Z bosons are much less frequently produced compared to Higgs for the reason we have discussed previously;  on the other hand, W bosons are much more frequently produced. In addition, since gluinos are flavor blind, they can decay to third-generation squarks. This leads to top quarks being produced in the cascade decay and further contributes to the large multiplicity of W bosons in events in this benchmark model.  

Benchmark 2 features a very small $\mu$-term, corresponding to a Higgs which is produced in the big cascade.  
This model is the same as SHSP 1a in \cite{Kribs:2010hp}. Due to large gluino mass, the production is dominated by the process $pp\rightarrow \tilde q\tilde q'$. This also indicates that few top quarks will be present in these events. In contrast to Benchmark 1, Z bosons are almost equally produced in the cascade decay as Higgses.

\begin{table}[htbp]
\begin{centering}
   \begin{tabular}{|l|c|c|}
   \hline
     Model & $1$ & $2$     \\ \hline
    $m_{\tilde q_{L,R}}$ &  $940,910$ & $1000$  \\
    $m_{\tilde\ell }$  &  1000  & 1000\\
    $m_{\tilde g}$ &  $949$ & $2036$  \\
    $m_{\chi_1^0}$ &  $163$ & $138$ \\ 
    $m_{\chi_2^0}$ &  $306$ & $-158$  \\ 
    $m_{\chi_3^0}$ &  $-518$ & $306$  \\ 
    $m_{\chi_4^0}$ &  $535$ & $625$ \\ 
    $m_{\chi_1^{\pm}}$ &  $305$ & $148$  \\ 
    $m_{\chi_2^{\pm}}$ &  $534$ & $625$ \\ 
    $\tan\beta$ &  $10$ & $10$ \\ 
    $\mu$ &  $512$ & $150$ \\ \hline
    $\sigma(\tilde g,\tilde q)$ & $2.5$~pb & $0.41$~pb  \\
    $\text{BR}(\tilde q_{L}\rightarrow h)$ & $30\%$ &  $22\%$ \\  
    $\text{BR}(\tilde q_{L}\rightarrow Z)$ & $3\%$& $25\%$  \\  
    $\text{BR}(\tilde q_{L}\rightarrow W)$ & $64\%$& $48\%$ \\ 
    $\sigma \cdot \text{BR}(h)$ & $0.29$~pb & $0.04$~pb \\ 
    $\sigma \cdot \text{BR}(h+W/Z)$ & $0.47$~pb & $0.1$~pb \\ 
    $\sigma \cdot \text{BR}(W/Z)$ & $1.04$~pb & $0.23$~pb \\ 
    \hline
  \end{tabular}
  \caption{The relevant masses, cross sections and branching ratios for the benchmark SUSY models. The spectrum and decay branching ratios were calculated using SUSY-HIT~\cite{Djouadi:2006bz}. $\sigma(\tilde g,\tilde q)$ are the $2\rightarrow 2$ LO cross sections involving $\tilde g$ and $\tilde q$, which were calculated in Pythia. 
 $\text{BR}(h)$, $\text{BR}(h+W/Z)$ and $\text{BR}(W/Z)$ are the branching ratios for events with at least one Higgs boson but no W/Z boson, with both Higgs and W/Z bosons, and with at least one W/Z boson but no Higgs boson respectively. Masses are given in GeV.}
  \label{table:mass}
\end{centering}
\end{table}

These two scenarios are, in a sense, \textit{orthogonal} to each other and capture effectively a large portion of the parameter space where the relevant branching ratios, barring accidental degeneracies in the spectrum, may vary only by a factor of 2-3 with respect to the ones appearing in Tab.~\ref{table:mass}.

\section{Simulation and SUSY Cuts}
\label{SUSYcuts}

Events are generated, showered, and hadronized in Pythia 6.4.24 ~\cite{Sjostrand:2006za} utilizing the ``DW tune"~\cite{Albrow:2006rt}. 
Initial and final state radiation as well as multi-particle interactions are turned on. We included pile-up events assuming a luminosity per bunch crossing of $0.05\text{ mb}^{-1}$. 
Supersymmetric events are isolated by means of cuts which render SM backgrounds negligible compared to the SUSY signal.  We use the cuts in SUSY analyses designed by the CMS collaboration for early LHC SUSY searches \cite{Hubisz:2008gg,Ball:2007zza}. 
First we require missing transverse energy $E^{\text{miss}}_T > 200$~GeV and at least three jets with $p_T > 30$~GeV with
pseudorapidity $|\eta| < 3$. 
In addition, only events in which the hardest and the second hardest jets have $p_{T} > 180,110$~GeV respectively are kept. 
The hardest jet is also required to be within the central tracker fiducial volume, i.e. $|\eta| < 1.7$. 
Finally, we require $H_T> 500$~GeV, where
$H_T = \sum_{i=2}^{4} p_{T}^{i} + E_{T}^{\text{miss}}$.
The jets in the above cuts are raw jets clustered using Cambridge/Aachen (C/A) algorithm~\cite{Dokshitzer:1997in} with a cone-size $R=0.5$ on calorimeter cells with granularity $\delta \eta\times \delta \phi = 0.1\times 0.1$ between $ -3 < \eta < 3$. 
The jet clustering in our analysis is performed using the FastJet(v-2.4.2)~\cite{Cacciari:2005hq} libraries. 

The preselection cuts and their associated cumulative efficiencies on the SUSY signal events from models $1$ and $2$ can be seen in Tab.~\ref{table:cuts}.  While relatively robust on the SUSY signal events, these cuts are far out on the tails of SM QCD, di-boson, and t-tbar backgrounds.  In the remaining sample, the primary obstruction to reconstructing a hadronically decaying Higgs are the SUSY events which include $W$'s and $Z$'s.  To reduce this background, we turn to more sophisticated jet substructure algorithms.
\begin{table}[htbp]
\begin{centering}
   \begin{tabular}{l|c|c}
    \hline
           cut/sample & 1 & 2 \\ \hline
           $E_T^{\text{miss}}>200$\text{GeV}& $80.64\%$ & $80.54\%$ \\ \hline
           $N_j \ge 3$ & $75.32\%$ & $78.87\%$ \\ \hline
           $p_{T,1}>180, p_{T,2} > 110$  & $72.29\%$ & $77.72\%$ \\ \hline
           $H_T>500$~GeV & $35.54\%$  & $54.47\%$ \\
    \hline
  \end{tabular}
  \caption{Cumulative efficiencies for the preselection cuts to isolate SUSY events.}
  \label{table:cuts}
\end{centering}
\end{table}

\section{Substructure analysis}
\label{substructure}

Even in the absence of backgrounds, reconstructing the Higgs boson from hadronic jets is generally difficult at hadron colliders. 
However, in supersymmetric production, the Higgs can easily get boosted which collimates all of the Higgs decay products into a ``fat jet." 
An effective algorithm for identifying these jets and reconstructing Higgs candidates is provided by Butterworth, Davison, Rubin and Salam (BDRS) \cite{Butterworth:2008iy}, where it was used for reconstructing SM Higgs.

This method starts by forming fat jets with cone size large enough to capture most of the hadronic products of the boosted Higgs; subsequently, one 
scans within this fat jet, looking for a particular jet substructure which corresponds to the presumed decay topology of the Higgs (in our analysis,  $h \rightarrow 2 \eta \rightarrow 4j$).
We describe the BDRS procedure in detail below:
\begin{description}
\item[(a)] Cluster hadronic calorimeter activity into jets by iteratively recombining pairs of closest distance $d_{ij}$. For C/A jet algorithm, $d_{ij}$ is given by the angular distance $\Delta R_{ij}\equiv \sqrt{ (\phi_{i}-\phi_{j})^2 +  (\eta_{i}-\eta_{j})^2}$. The recombination ends when all objects are separated by some minimum $\Delta R_{ij} > R$. Here $R$ is chosen to be large enough to contain the decay products of the Higgs. 
\item[(b)] Uncluster each fat jet into two subjets $j_1$ and $j_2$ ($m_{j1}> m_{j2}$).  
Two criteria must be satisfied by these subjets in order to associate the fat jet to some presumed heavy parent particle.  
First,  there must be a significant mass drop $m_{j1} <  \mu \; m_j$, where $m_j$ is the total invariant mass of the parent fat jet and $\mu$ is a cut parameter.  
Second, it is required that there is no significant asymmetry in the two subjets defined by:  $y\equiv {\rm min}(p_{T\,j_1}^2,p_{T\,j_2}^2) / m_j^2 \;\Delta R^2_{j_1,j_2} > y_{cut}$. 
When these two conditions are satisfied we exit the loop and dub the jet as candidate Higgs jet.
\item[(c)] If the subjets do not satisfy the above requirements, then $j_1$ is identified as a new fat jet, and step (b) is repeated by subdividing $j_1$ into a sub-jet pair.  This is repeated until either $p_{T,j_1}<50$~GeV or $j_1$ can no longer be unclustered, at which point the initial fat jet is discarded as a candidate for a massive parent. 
\end{description}
In our analyses, we take $R=0.9-1.2$, $\mu = 0.5 - 0.667$ and $y_{\text{cut}}=(0.3)^2$, with small variations for different situations specified in Tab.~\ref{table:subjetcut}.  
Note, that the BDRS algorithm is no very effective at selecting the Higgs events: only about 10\% of the signal survives the initial subject analysis. This is due to the fact that the boost of the higgs boson is not that incredibly large in these cases. 

\begin{figure}[htbp]
\begin{center}
   \includegraphics[width=3.0in]{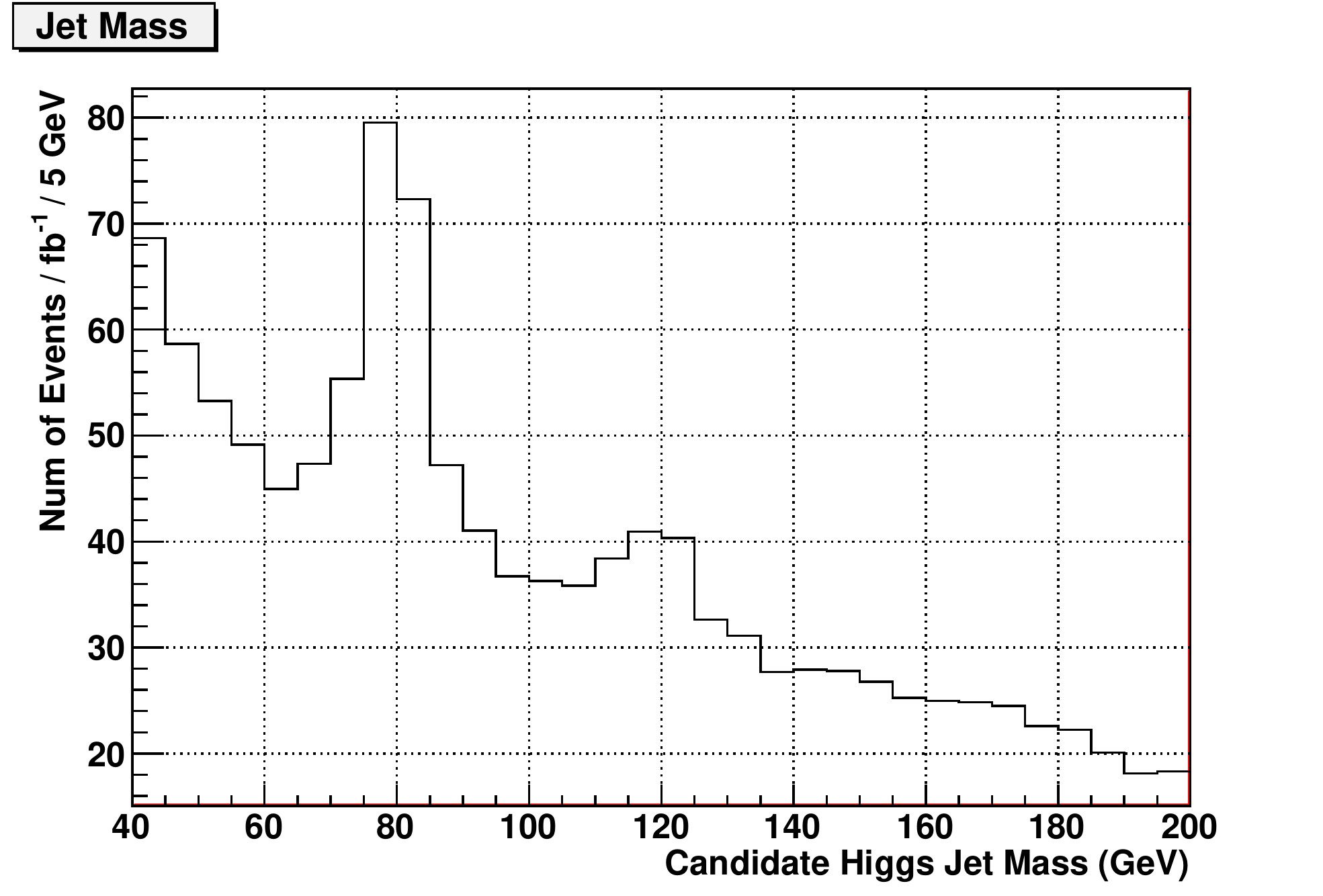}
\caption{The candidate Higgs jet mass distribution for SUSY benchmark 1 obtained using the BDRS algorithm.  The Higgs mass and the $\eta$ mass are $m_h=120$~GeV and $m_{\eta}=10$~GeV respectively.}
\label{susy1_OnlyBDRS}
\end{center} 
\end{figure}

Compared to the SM Higgs events, the supersymmetric events are typically of much greater multiplicity, often containing multiple electroweak gauge bosons in addition to hard quark and or gluon jets from the decay of squarks and gluinos.
 Many of these hard jets and hadronically decaying $W$'s and $Z$'s  can be misidentified as Higgs candidates, even after the substructure analysis is performed.  This is the primary obstruction to discovering the Higgs as a peak in the jet mass distribution.
 The issue is compounded by the fact that the $W$ and $Z$ bosons lie close to the mass range expected for the Higgs. 
Fig.~\ref{susy1_OnlyBDRS} shows the typical jet mass distribution obtained for the candidate Higgs event. 
We therefore must supplement the BDRS algorithm with additional cuts to suppress the $W$ and $Z$ contamination of the Higgs sample as well as other combinatoric jet backgrounds. Different strategies are necessary for different $\eta$ masses.  We discuss them in sequence below.

For the low $m_{\eta}$ case, the decay products from the $\eta$ decay are collimated, and therefore the jet substructure is close to the SM Higgs case $h\rightarrow b\bar{b}$. 
In this case, the two subjets from unclustering the fat jet are expected to correspond the two $\eta$-jets. In order to reduce the contamination from W/Z jets, one could consider additional cuts on the following variables as discussed in \cite{Falkowski:2010hi}:
\begin{description}
\item[Mass democracy:]  $$\alpha_{\text{MD}} \equiv \frac{\text{min}(m_{j1},m_{j2})}{\text{max}(m_{j1},m_{j2})}$$
\item[Flow variable:]  $$\beta_{\text{flow}} \equiv \frac{p_{T, j3}}{p_{T, j1}+p_{T, j2}}, \;\; \mbox{ if }\;\; p_{T,j3} > p_{T}^{min}.$$
\end{description}
For Higgs decay through two light $\eta$'s, we expect $\alpha_{\text{MD}}\sim 1$ and $\beta_{\text{flow}}\ll 1$. This is based on the fact that both higgs and $\eta$ are QCD singlets and therefore  radiation only occurs at the virtuality scale $\sim m_{\eta}$ after the $\eta$ has decayed. The reduced radiation indicates small $\beta_{\text{flow}}$ and also small shift in the $\eta$ jet mass. This is in contrast to the QCD jets, where the virtuality scale is governed by the initial hard scattering. In \cite{Falkowski:2010hi}, cuts on these variables were used to separate the Higgs jet from the QCD jet.  In our case, they can instead reduce the combinatoric jet backgrounds that are present together with the Higgs. In addition, the mass democracy and the flow variable cuts are quite useful in distinguishing Higgs jets from W/Z jets since the final state radiation in $W/Z$ decay is at a larger scale $\sim m_{W/Z} \gg m_{\eta}$.  For example, in benchmark model 1 with $m_{h}=120$ GeV and $m_{\eta}=10$ GeV they cut roughly $75\%$ of the $W/Z$'s whereas $30\%$ of the Higgses.

For the high $m_{\eta}$ case, the decay products of $\eta$'s are less collimated while the two $\eta$'s are more collimated. This makes the four partons more uniformly distributed inside the fat jet, giving rise to a truly four-jet decay. This is most obvious in the low $m_{h}$ case, where the allowed phase space to decay into $\eta$ is limited.  In this case, the two subjets found by unclustering the fat jet may not match the partonic object from one of the $\eta$'s. In addition, due to the increased multiplicity of the decay, the subjets are typically softer.  In order to reduce the W/Z background, we need different cuts compared to the light $\eta$ case.
\begin{description}
\item[Number of subjets:]  The simplest option is to require at least four hard subjets inside the fat jet obtained from the BDRS procedure: we re-cluster the candidate fat jet  into $n_{\text{subjet}}$ subjets with a smaller cone size $R_{\text{sub}}$,
  $$n_{\text{subjet}}\ge 4 \quad\mbox{ with } \quad p_T>15\mbox{ GeV.}$$
\end{description}
This is easy to understand since W/Z jets typically only have two hard subjets.

Another possibility is to use the planar flow variable introduced in \cite{Almeida:2008yp}, which is sensitive to whether the color flow is linear or isotropic. The planar flow vanishes for linear shapes and approaches unity for isotropic depositions of energies. In the context of $h\rightarrow 2\eta\rightarrow 4j$, the planar flow increases as $m_{\eta}$ increases since the final states become more isotropic. However, in the cases that we studied, the number-of-subjet cut is already very effective, and we do not include the planar flow in our final result.

We have also investigated whether the jet pull variable~\cite{Gallicchio:2010sw}
significantly enhances signal relative to background.  We found that, in the cases we analyzed, there is little to no improvement as the signal distribution in this variable is too similar to the SM gauge boson background. 
However, we have not performed a multivariable combined study that could partially enhance the significance~\cite{Black:2010dq}.

\begin{figure}[htbp]
\begin{center}
   \includegraphics[width=3.0in]{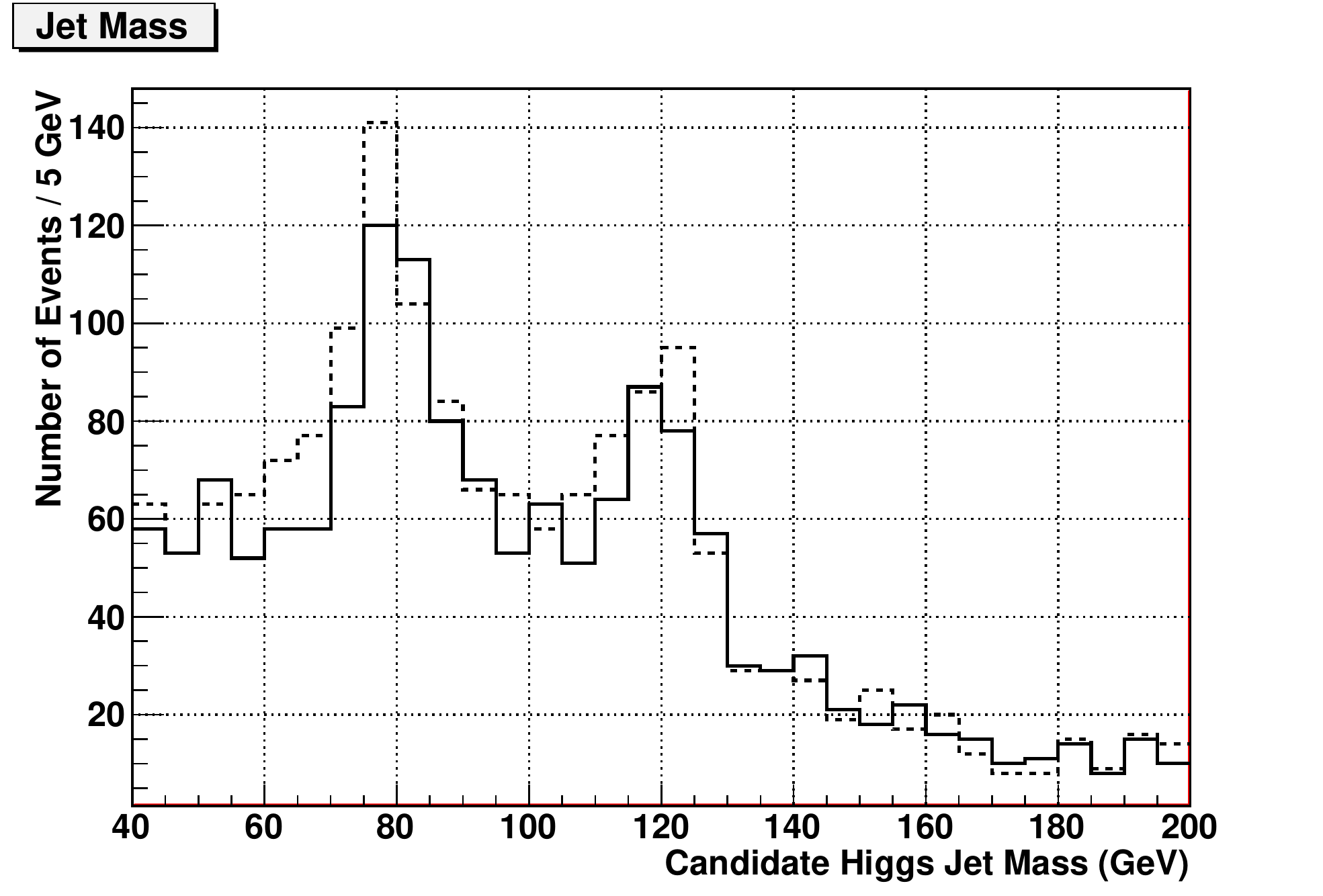}
   \includegraphics[width=3.0in]{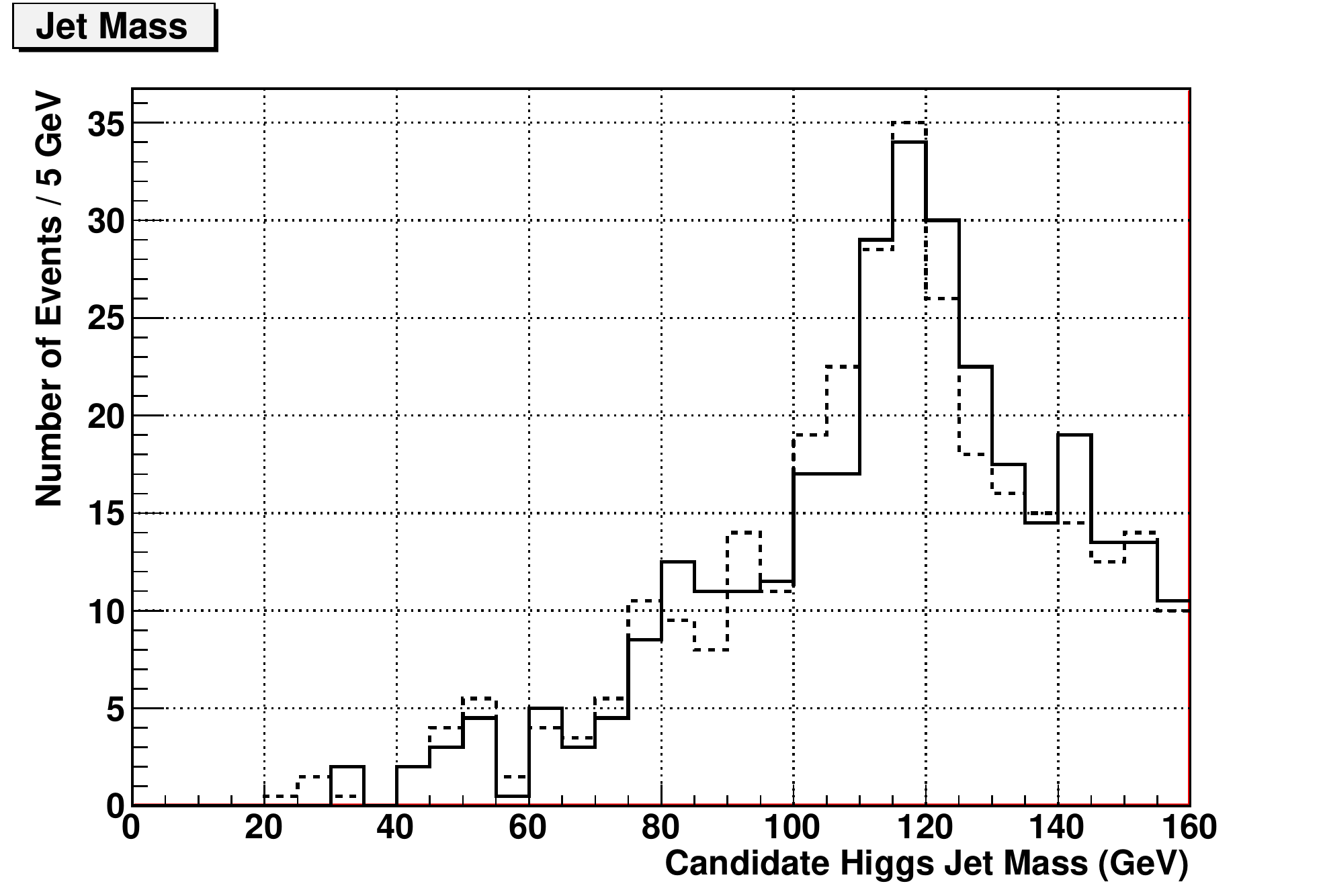}
\caption{The candidate Higgs jet mass distribution for SUSY benchmark 1 with (solid) and without (dash) pile-up events. Top: The Higgs mass and the $\eta$ mass are $m_{h}=120$~GeV and $m_{\eta}=10$~GeV respectively. Bottom: $m_{h}=120$~GeV and $m_{\eta}=30$~GeV.
The plots are generated using  $10\mbox{ fb}^{-1}$. Substructure cuts are given in Table \ref{table:subjetcut}.}
\label{Plot_Pile-up}
\end{center}
\end{figure}

In the last step of the reconstruction, a filtration algorithm cleans up these candidate jets by removing soft components. 
For low $\eta$ mass, one decomposes the fat jet to subjets by taking a smaller $R_{\text{sub}}$, 
and sum up the leading $n_{\text{filt}}$ subjets to obtain the filtered jets.
In our analysis, we take $R_{\text{sub}}= \text{min}(\Delta R_{j1,j2}/2, 0.3)$.
For high $\eta$ mass, we \textit{trim} it by only keeping subjets with $p_T > f_{\text{cut}}\,  p_{T,J}$~\cite{Krohn:2009th}, where $R_{sub}=0.2-0.3$. It should be noted that the threshold $f_{\text{cut}}$ affects both the accuracy and resolution of the Higgs mass. 
For smaller threshold $f_{\text{cut}}$, meaning more decay products of the Higgs would be included, the reconstructed Higgs mass would be closer to the true mass. On the other hand, it is also easier for the contamination from other softer partons in the same event to leak into the Higgs jet, which would worsen the mass resolution. 
The effects of pile-up events can be seen from e.g. Fig.~\ref{Plot_Pile-up}. For light $m_{\eta}$, with pile-up events included, it is harder for the fat jet to pass the flow cut. This leads to a decrease in the W and Higgs peaks. But on the other hand, the continuum background also drops. 
For heavy $m_{\eta}$, there are no qualitative changes in the candidate jet mass distribution. 
For convenience, we present our final result in the figures of Sec.~\ref{results} without pile-up events. 

Tab.~\ref{table:subjetcut} shows the substructure cuts that we use in our search. We will discuss the details in the next section. 
The concrete values of the cuts vary case by case depending on the Higgs and the $\eta$ mass, but they are not optimized yet.
\begin{table}[htbp]
\begin{centering}
   \begin{tabular}{lcccr}
    \hline
      $m_{h},m_{\eta}$  & $\;\;(120,10)\;\;$ & $\;\;(100,10)\;\;$     & $\;\;(120,30)\;\;$  & $(100,30)$ \\ \hline
             $R$                & $1.2$          & $1.2$      & $1.0$   & $0.9$ \\ \hline
             $\mu$            & $0.667$       & $0.667$   & $0.667$   & $0.5$  \\ \hline 
 $\alpha_{\text{MD}}$  & $>0.7$        &  $>0.8$   & $>0.4$ & $>0.4$   \\ \hline
 $\beta_{\text{flow}}$   & $<2\%$       &  $<0.5\%$  &  -           & -         \\ 
 $p_T^{\text{min}}$     & $2.0$           &   $1.0$       &  -           & -         \\  \hline
   $R_{\text{sub}}$       & -                  &  -            & $0.25$    & $0.25$ \\ 
   $n_{\text{subjet}}$    & -                  &  -             & $\ge 4$   & $\ge 4$ \\
 $p_{T,\text{sub}}^{\text{min}}$  & -              &  -            & $15$    & $17$ \\ \hline
  \end{tabular}
  \caption{Jet substructure cuts for different scenarios. Momenta and masses are in unit of GeV.
The cuts in the column $(100,10)$ are only for benchmark 2. }
  \label{table:subjetcut}
\end{centering}
\end{table}

\section{Results}
\label{results}

We now apply this method of Higgs reconstruction to the two SUSY benchmark models for different Higgs and $\eta$ masses.  The Higgs appears as a resonance peak in the jet-mass distribution of the fat jets which survive the substructure cuts.  While the substructure analysis is reasonably successful at removing hadronically decaying $W$ and $Z$ bosons, significant contamination of the sample in the $80-90$~GeV region from these resonances remains.  The low Higgs mass region, where LEP could have missed the Higgs thus remains especially challenging.  We consider separately two different Higgs mass regions: high mass ($m_h \gtrsim 115$~GeV) and low mass ($m_h \lesssim 100$~GeV). 

In the heavy Higgs mass region, there is little interference from $W$ and $Z$ contamination of the fat jet sample since the peaks in the jet mass distribution are well separated.  
In this case, one does not need to completely suppress the contribution from hadronically decaying $W$'s and $Z$'s, and lower luminosity will be sufficient for Higgs discovery. 

In the low mass region, the $W$ and $Z$ jet mass peaks share significant overlap with a potential Higgs signal, unless the contamination of $W$ and $Z$ bosons can be significantly reduced without losing too much of the Higgs signal efficiency.  This is in principle possible, due to the different decay topology of these events, although issues arise when the $\eta$ is too light.

In the case of light Higgs and heavier $\eta$, the 4 subjets arising from the two $\eta$ decays are often resolvable.  Additional cuts on the number of sub-jets appearing within the fat jet are therefore effective at removing $W$'s and $Z$'s, even for a relatively light Higgs boson.  In the benchmark models we consider, the $W$ and $Z$ background is low enough to identify the Higgs.

For the scenario of both light $\eta$ and Higgs mass below $100$~GeV, we find that we cannot remove a large enough fraction of the $W$ and $Z$ boson events to be assured that an excess in this mass range is due to a Higgs.  This is due to the fact that light $\eta$'s will have substantial relativistic boost and correspondingly collinear decay products.  The Higgs decay then appears to have di-jet substructure, just like the SM gauge bosons.  Substructure cuts therefore reduce both signal and background to a similar degree.  

One approach to remedy this could be to try to understand the details of the SUSY background and subtract it.  This could be done for the $Z$ boson, for example, by measuring the number of reconstructed $Z$ bosons in the leptonic decay channel.   Unfortunately, this can not be done with the $W$ boson, since the semi-leptonic decay involves a neutrino whose momenta is lost along the LSP contribution to the total missing $p_T$.  Even armed with perfect knowledge of these rates, such subtractions are especially susceptible to systematic uncertainties in the shape of the $W$ and $Z$ jet mass distributions.  Due to these difficulties, we do not attempt such a subtraction.

To illustrate the effectiveness of the substructure cuts, we looked into our data sample and identified the associated heavy object for a given candidate fat jet. The associated object is defined to be the closest heavy object within an $R=0.4$ cone around the jet. Given that information, we are able to count the number of ``correct" Higgs jets in $\pm 5$~GeV window around the true mass, and the total number of candidate jets in that window. Similarly this can be done for W and Z bosons. These numbers can be compared with the number of Higgs, W or Z bosons in the sample without subjet cuts to get an estimate of the efficiency and the discrimination power. As can be seen in Tab.~\ref{table:cutseff}, a factor of $\sim 20$ gain in efficiency can be achieved for Higgs against W and Z for $m_{\eta}=30$~GeV, while a factor of $\sim 5$ for $m_{\eta}=10$~GeV.

\renewcommand{\arraystretch}{1.35}
\begin{table}[htbp]
\begin{centering}
\begin{tabular}{l|c|c|c|c|c|c|c|c}
\hline
&\multicolumn{4}{c|}{Model 1}&\multicolumn{4}{c}{Model 2}
\\ \hline
&\multicolumn{2}{c|}{$(100,10)$}&\multicolumn{2}{c|}{$(100,30)$}
&\multicolumn{2}{c|}{$(100,10)$}&\multicolumn{2}{c}{$(100,30)$}
\\ \hline
&\multicolumn{1}{l|}{before}&\multicolumn{1}{l|}{after}
&\multicolumn{1}{l|}{before}&\multicolumn{1}{l|}{after}
&\multicolumn{1}{l|}{before}&\multicolumn{1}{l|}{after}
&\multicolumn{1}{l|}{before}&\multicolumn{1}{l}{after} 
\\
 \hline
H & 6974  & $\frac{324}{473}$ &  6587     & $\frac{69}{103}$ & 22450 & $\frac{700}{831}$  & 22564 & $\frac{298}{403}$
\\ \hline
W      & 22668 & $\frac{366}{581}$ &  22435  & $\frac{7}{26}$   & 63641  & $\frac{356}{564}$   & 62775 & $\frac{34}{274}$ 
\\ \hline
Z       & 1296   &  $\frac{18}{390}$  &  1244    & $\frac{0}{67}$   & 22977  & $\frac{136}{671}$   & 22933 & $\frac{19}{269}$ 
\\ \hline
\end{tabular}
  \caption{Subjet cut efficiencies for Higgs and W/Z bosons in the window $\pm 5$~GeV around their true masses. The number before cuts are the number of Higgs, W or Z in the event sample after applying preselection cuts on the $10^5$ raw events. The number after cuts is presented in the form $\frac{a}{b}$ where $a$ is the number of ``correct" Higgs, W or Z jets and $b$ is the total number of candidate jets in the respective mass window. The ``correct" Higgs, W or Z jets are defined as those candidate jets where the closest heavy object within $R=0.4$ cone around the jet is Higgs, W or Z. }
  \label{table:cutseff}
\end{centering}
\end{table}
\renewcommand{\arraystretch}{1}

\subsection{Low $\eta$ Mass ($m_{\eta} =10$~GeV)}
\label{LowMass}
 
For the low $m_{\eta}$ case, we use the modified BDRS method with mass democracy cuts and flow cuts to identify Higgs jets.  As discussed above, the substructure analysis is not be able to substantially reduce the contribution of $W$ and $Z$ bosons while preserving Higgs signal events.  For both benchmark models with $m_h = 120,100$~GeV, we find candidate Higgs jets and construct the jet-mass distribution.

\begin{figure}[htbp]
\begin{center}
   \includegraphics[width=3.in]{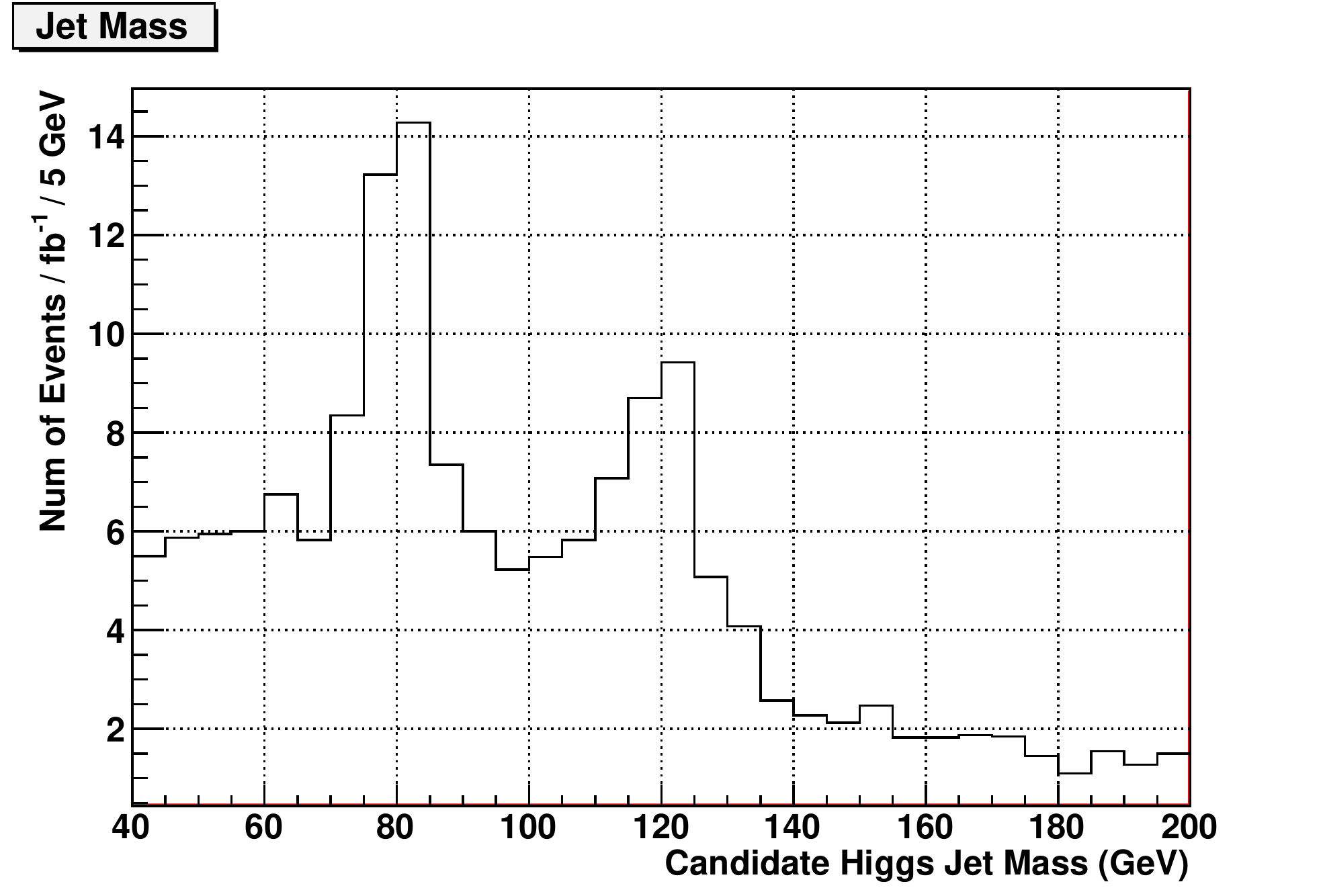}
   \includegraphics[width=3.in]{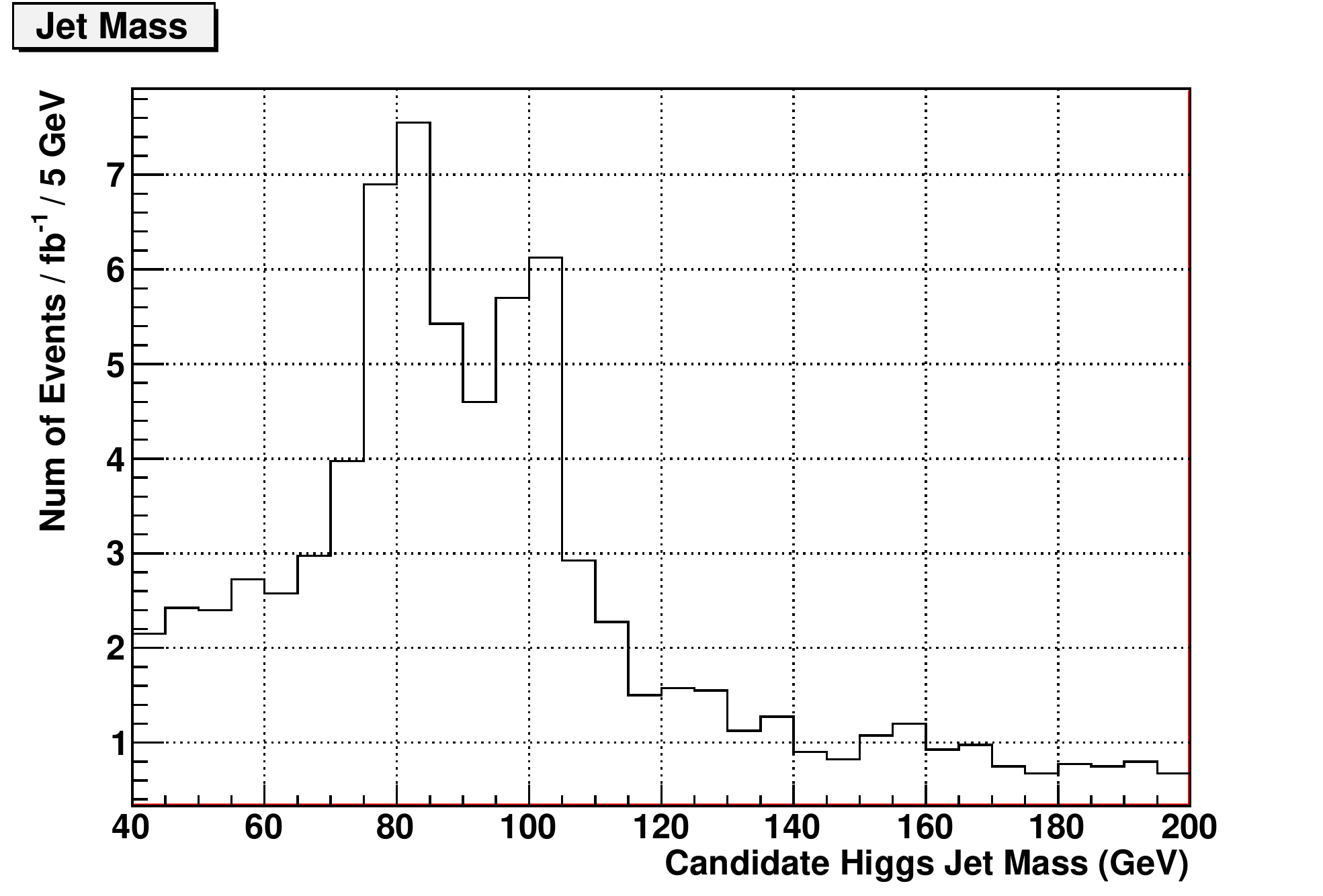}
\caption{The candidate Higgs jet mass distribution for SUSY benchmark 1. Top: $m_h=120$~GeV and $m_{\eta}=10$~GeV. 
Bottom: $m_h=100$~GeV and $m_{\eta}=10$~GeV. 
Events with $\ge 7$ jets ($p_T> 30$~GeV) are vetoed in the bottom plot.}
\label{higgs-mass:susy1-low-eta}
\end{center}
\end{figure}

For benchmark model 1, we take the  values for the cut parameters to be $R=1.2$, $\alpha_{\text{MD}}>0.7$, $\beta_{\text{flow}} < 2\%$ and $n_{\text{filt}}=3$.  The results for $100,000$ raw events normalized by the cross section are shown in Figure~\ref{higgs-mass:susy1-low-eta} for both high Higgs mass (top panel) and low  Higgs mass (bottom panel).  In this plot, the Higgs mass peaks are well above the background and its position is consistent with the true Higgs mass.  The peaks in the vicinity of $80$~GeV are from hadronically decaying $W$'s which evade the above cuts. To calculate the significance of the Higgs peak, we must provide an estimate of the backgrounds from both SM and SUSY. The SM backgrounds are negligible as we discussed before and are taken to be zero for simplicity, while the SUSY backgrounds can be estimated from the continuum under the Higgs peak in the jet-mass distribution. For example, for the case with $m_h=120$~GeV, we take the $-2$/$+1$ bins around the peak $120$~GeV as the signal region and the two adjacent bins for background estimation. We find that a $5\sigma$ discovery of the Higgs boson for $\sim 10\,\text{fb}^{-1}$ is possible.  For the case of low Higgs mass, in the bottom panel of this Figure, these two mass peaks are closer. Taking the excess in the $\pm 1$ bins around the peak $100$~GeV as the signal, a $5\sigma$ significance can also be achieved with the same amount of data. For an even smaller Higgs mass, the signal peak would begin to merge with the $W$ peak.  Unless the $W$ fake rate can be further reduced with additional novel techniques, it seems unlikely that a Higgs with mass much smaller than $100$~GeV can be identified.

\begin{figure}[htbp]
\begin{center}
   \includegraphics[width=3.in]{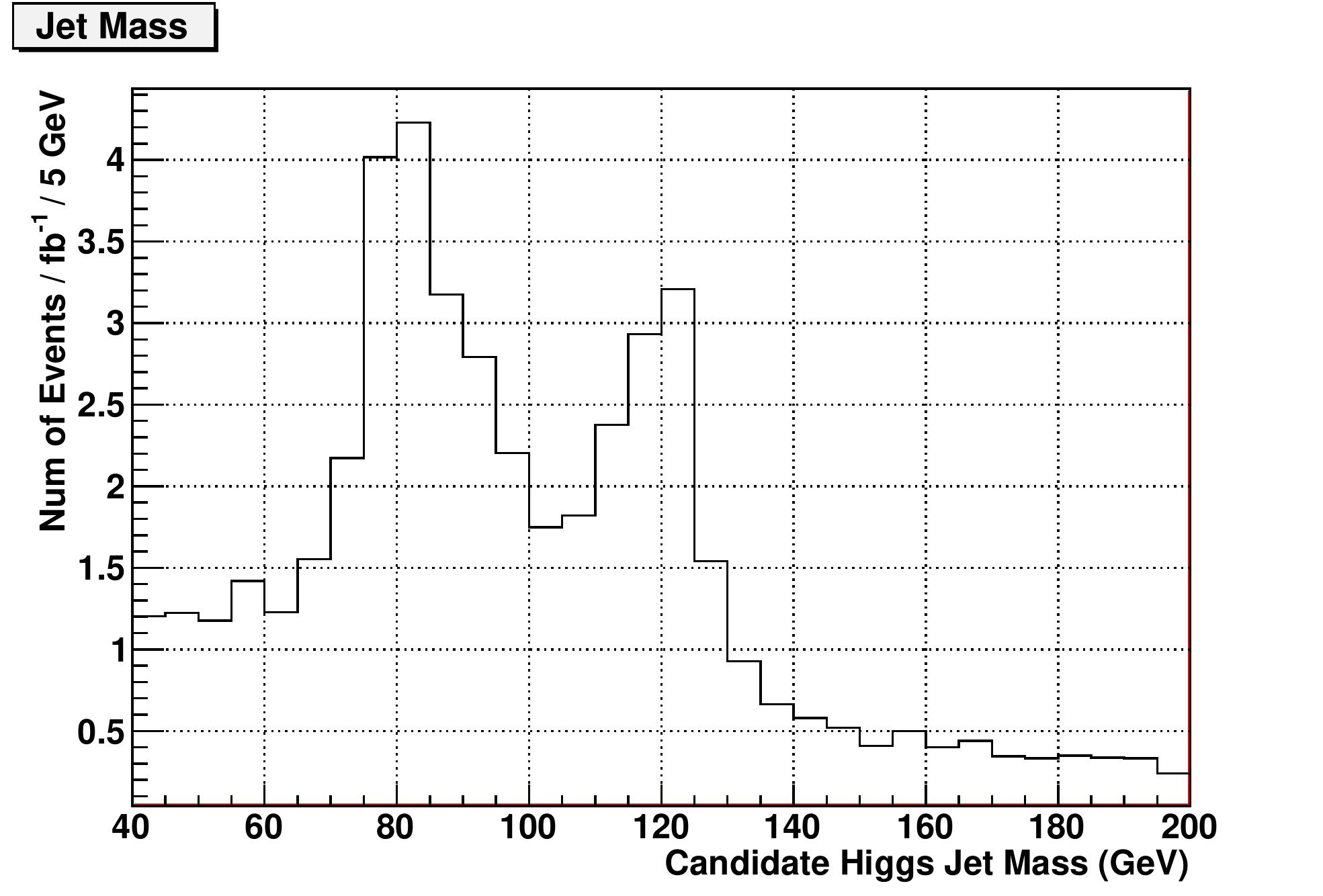}
   \includegraphics[width=3.in]{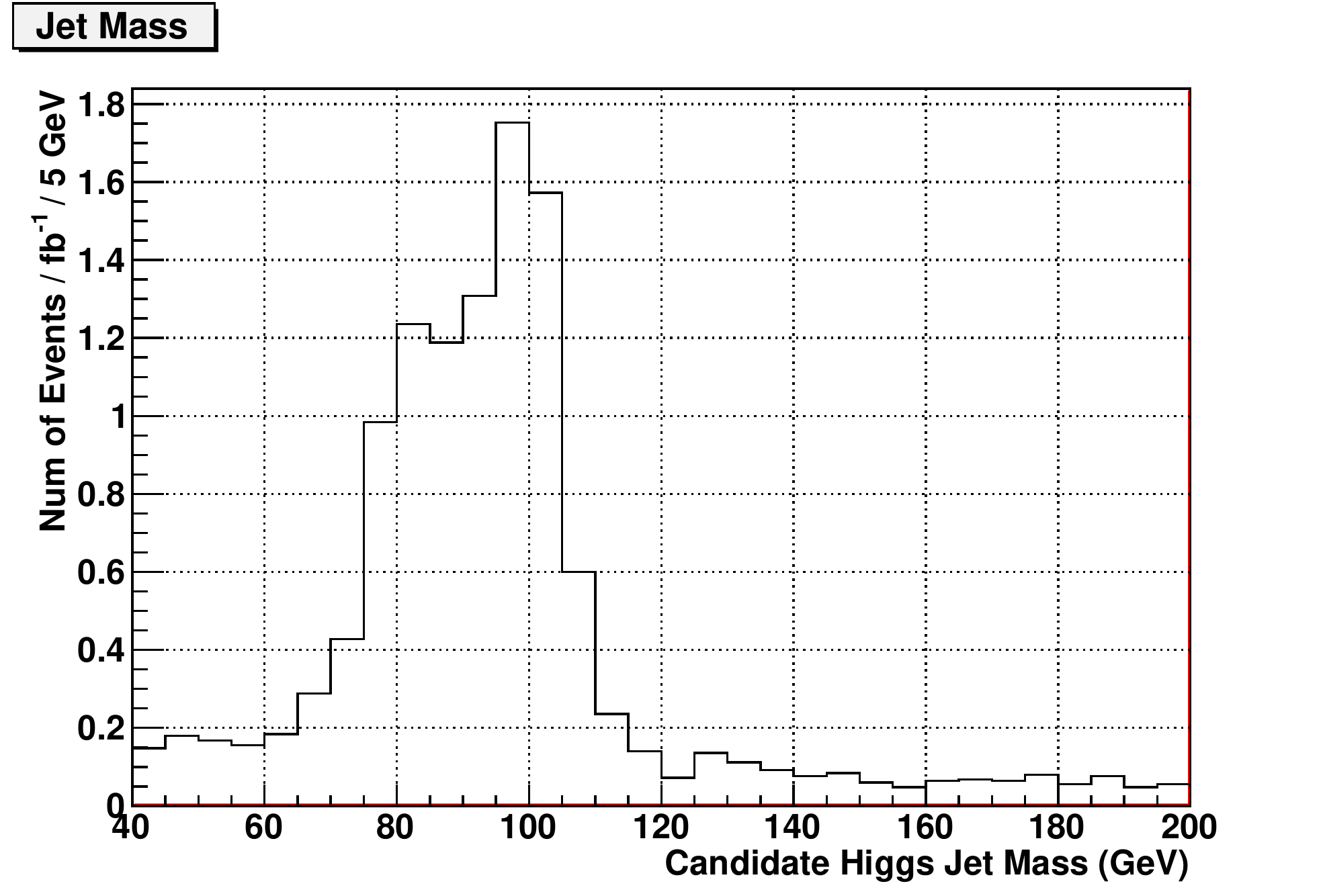}
\caption{The candidate Higgs jet mass distribution for SUSY benchmark 2. Top: $m_h=120$~GeV and $m_{\eta}=10$~GeV. 
Bottom: $m_h=100$~GeV and $m_{\eta}=10$~GeV. 
}
\label{higgs-mass:susy2-low-eta}
\end{center}
\end{figure}

For benchmark model 2, 
the results are shown in Figure \ref{higgs-mass:susy2-low-eta}. In the top panel, we use the same cuts as for benchmark model 1 and we can see that a $5\sigma$ discovery can again be achieved (using $-2$/$+1$ bins for signal) for $\sim 10\,\text{fb}^{-1}$ integrated luminosity. In fact, in this case the Higgs bosons are generally more boosted due to the larger neutralino mass difference. This leads to a higher reconstruction efficiency than for benchmark 1, and even without the flow cuts we can obtain similar results with smaller luminosity. However, for the low Higgs mass, the distribution obtained from using the same cuts show a plateau between $80-100$~GeV. This is due to the superposition of W, Z and Higgs contributions. Imposing stronger cuts $\alpha_{\text{MD}}> 0.8$ and $\beta_{\text{flow}} < 0.5\%$ with $p_{T}^{\text{min}}=1$~GeV, lead us to the second plot in Figure \ref{higgs-mass:susy2-low-eta}.  While the W peak is now significantly suppressed, and the big peak located around $100$~GeV suggests the presence of the Higgs boson, the subtraction of the Z-background is needed in this case. Naively using the same prescription for calculating the significance, we find $5\sigma$ discovery can be achieved with $\sim 25\,\text{fb}^{-1}$ integrated luminosity.

\subsection{High $\eta$ Mass  ($m_{\eta} =30$~GeV)}
\label{HighMass}

Moving to the high $\eta$ mass case, the decays of the Higgs are more four-jet like. We use the BDRS algorithm supplemented with a cut on the number of subjets to find the Higgs-like jet. We re-cluster the candidate fat jet into subjets using $R_{\text{sub}}=0.25$ and require $n_{\text{subjet}} \ge 4$ hard subjets with $p_T>15$~GeV. The final candidate Higgs jets are obtained after trimming with threshold $f_{\text{cut}} = 1.5\%$. For the low-mass Higgs, the cuts are slightly adjusted as seen in Tab.~\ref{table:subjetcut}.

\begin{figure}[htbp]
\begin{center}
   \includegraphics[width=3.0in]{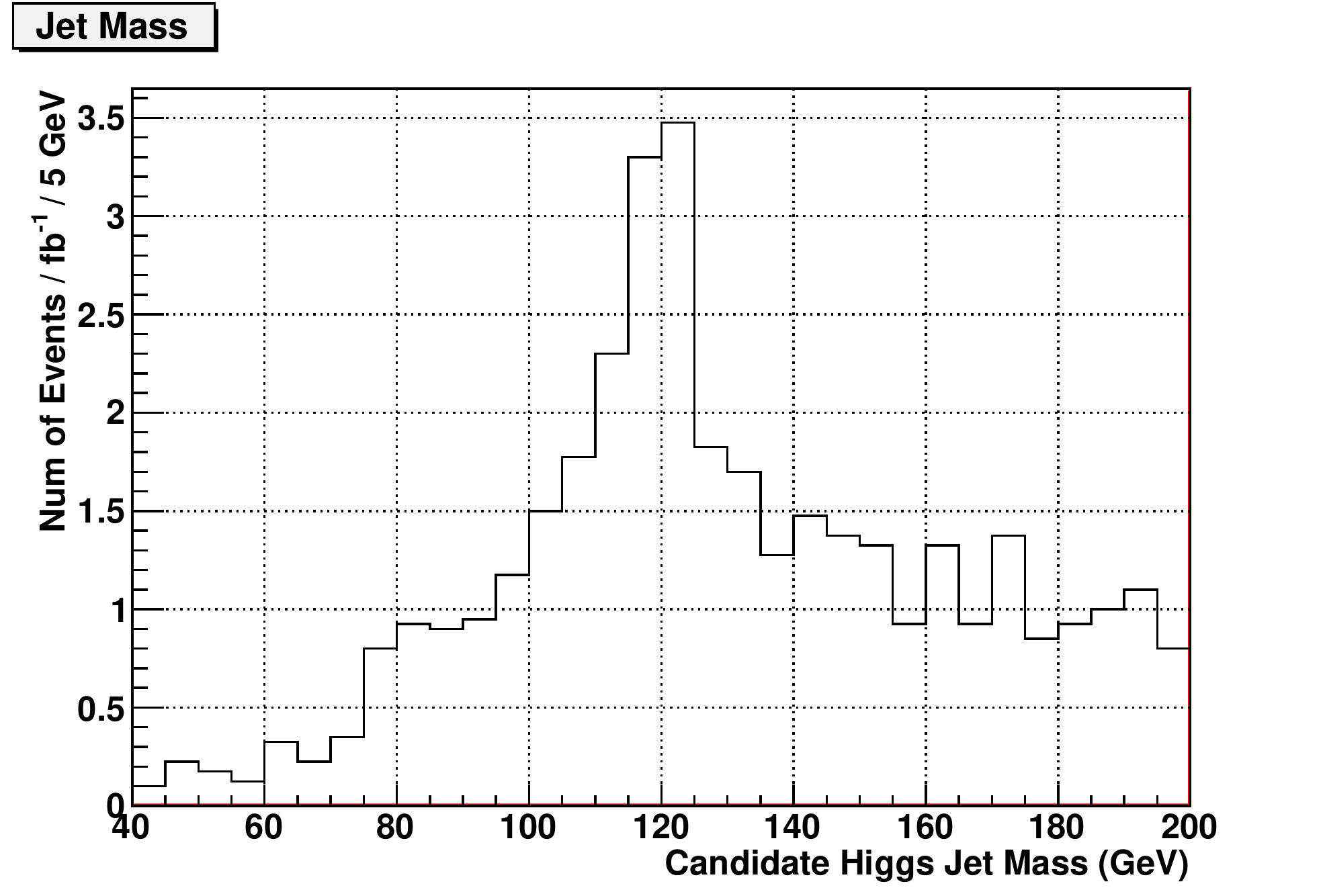}
   \includegraphics[width=3.0in]{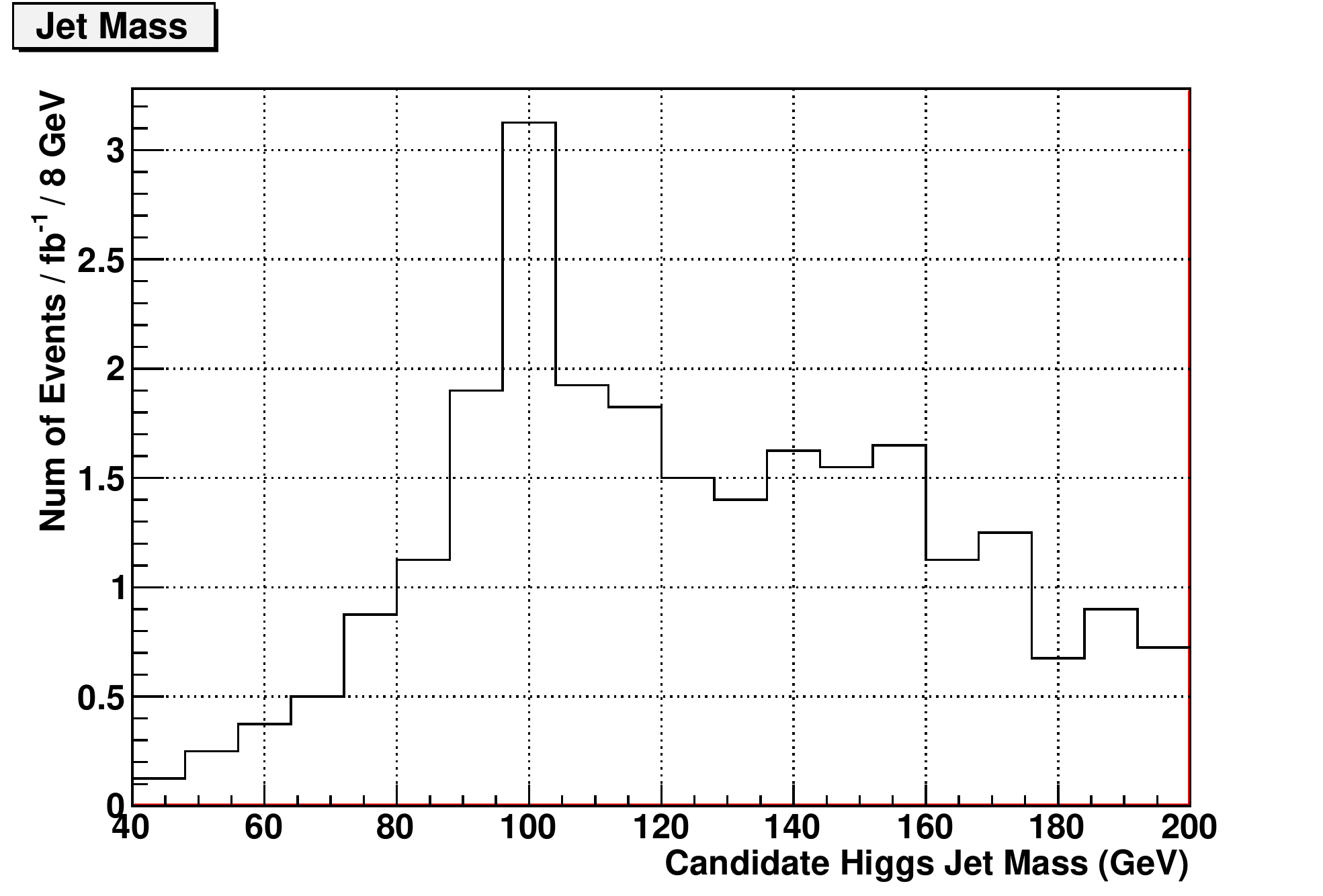}
\caption{Candidate Higgs jet mass distribution for SUSY benchmark 1. Top: $m_h=120$~GeV and $m_{\eta}=30$~GeV. 
Bottom: $m_h=100$~GeV and $m_{\eta}=30$~GeV.  Events with $\ge 8$ jets ($p_T> 30$~GeV) are vetoed in both plots. }
\label{higgs-mass:susy1-high-eta}
\end{center}
\end{figure}

\begin{figure}[htbp]
\begin{center}
   \includegraphics[width=3.0in]{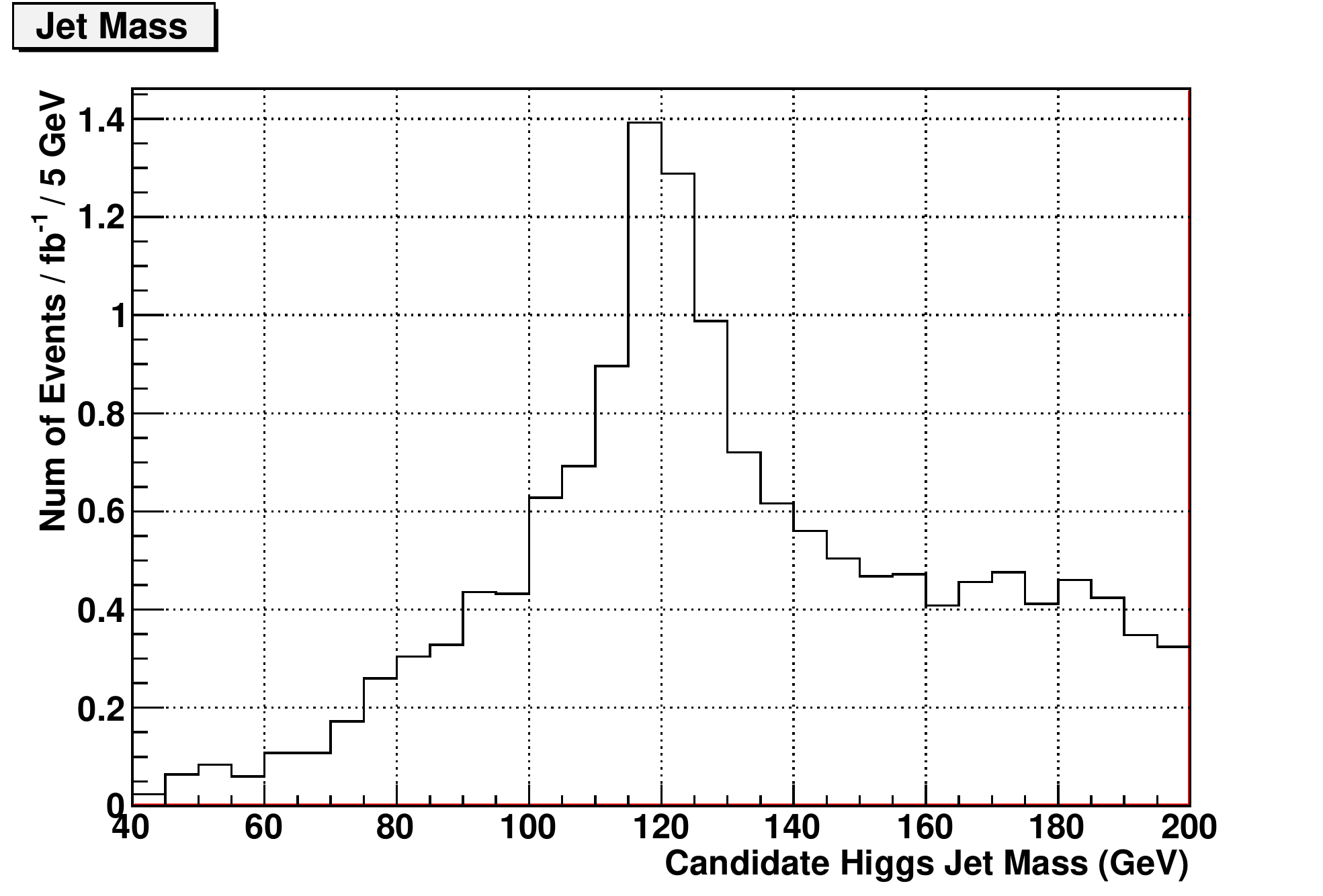}
   \includegraphics[width=3.0in]{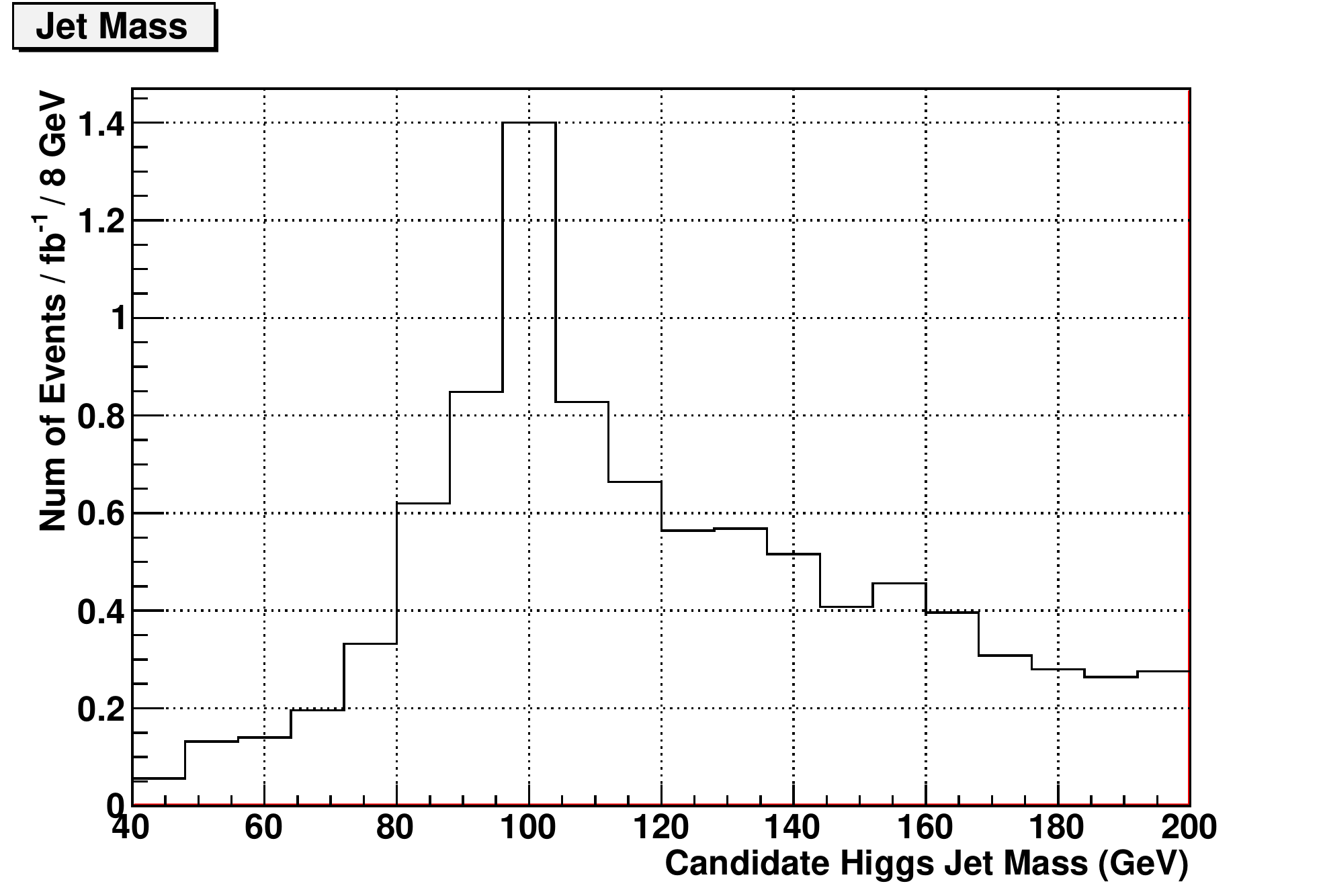}
\caption{Candidate Higgs jet mass distribution for SUSY benchmark 2. 
Top: $m_h=120$~GeV and $m_{\eta}=30$~GeV. 
Bottom: $m_h=100$~GeV and $m_{\eta}=30$~GeV. 
}
\label{higgs-mass:susy2-high-eta}
\end{center}
\end{figure}

The resulting candidate Higgs jet-mass distributions can be seen in Figures \ref{higgs-mass:susy1-high-eta},\ref{higgs-mass:susy2-high-eta}. Different from the low $\eta$ mass cases, the continuum background is small in the low mass region and the W/Z peaks are no longer visible. 
This indicates that the cut on the number of subjets is very efficient in reducing the W/Z contamination. But other combinatoric jet configurations can potentially leak through the cut since these may have more than two hard components and can give rise to a large jet mass. To suppress these combinatorics, we use a slightly smaller $R$ parameter for the jet clustering algorithm, and include a mild cut on the subjet mass democracy $\alpha_{\text{MD}}$ as shown in Tab.~\ref{table:subjetcut}. For benchmark 1, we require maximum $7$ jets in the events to further suppress the combinatoric background since there are lots of top quarks in the events.    
For the high Higgs mass case, the Higgs peaks are well reconstructed, and in both benchmarks a $5\sigma$ discovery can be achieved with roughly $10$ and $25\,\text{fb}^{-1}$ respectively (using $-2$/$+1$ and $\pm 2$ bins for signals). The results for low Higgs mass are similar, but more luminosity ($\gtrsim 35\,\text{fb}^{-1}$) is needed due to smaller signal efficiency. 


\section{Conclusions}
\label{conclusions}

Purely hadronic Higgs decays of the form $h \rightarrow 2\eta  \rightarrow 4j$ present a formidable, but not insurmountable challenge at the LHC.   Such Higgs bosons, when produced in association with massive exotica, can in many cases be reconstructed.  
In supersymmetric buried Higgs models, cuts on events with high $p_T$ jets and missing energy remove the QCD backgrounds that would otherwise swamp the Higgs resonance.  

Once QCD events have been stripped away, the remaining issue is to identify a Higgs resonance among the hadronic activity that occurs in typical SUSY cascade decay chains.  We have applied a substructure analysis that aids in the removal of remaining combinatoric backgrounds and/or dijet pairs arising from weak gauge boson decay.  

For buried Higgs bosons in the range $95-120$ GeV, with both heavy $(30~\text{GeV})$ and light $(10~\text{GeV})$ $\eta$ mass, discovery at the $5\sigma$ level is possible with $10$~fb$^{-1}$ of $7$ on $7$~TeV LHC running.  At lower values of the Higgs mass (and low $\eta$ mass), when the Higgs peak overlaps significantly with the $W$ and $Z$ resonances, substructure cuts do not sufficiently reduce weak gauge boson backgrounds.  In this case, discovery will be far more challenging, and will require more sophisticated techniques than those presented here.

\section*{Acknowledgements}
We would like to thank David Krohn for useful discussions and for comments on the manuscript. JH and JS thank Cornell University for their hospitality where part of the research was conducted. JS would also like to thank the Center for High Energy Physics at Peking University and Shanghai Jiaotong University for their hospitality where part of the research was conducted. This work of JS is supported by the Syracuse University College of Arts and Sciences. JH is supported by the Syracuse University College of Arts and Sciences, and by the U.S. Department of Energy under grant DE-FG02-85ER40237. This work of BB and CC has been supported in part by the NSF grant PHY-0757868.

\end{document}